\documentclass[aps,prb,showpacs,preprint,superscriptaddress]{revtex4}

\usepackage{latexsym} \usepackage{amsmath} \usepackage{dcolumn}
\usepackage{bm} \usepackage{graphicx} \usepackage{epsfig}
\graphicspath{{./Figures/}}

\begin{document}
%

\title{Optical to UV spectra and birefringence of SiO$_2$ and TiO$_2$: 
First-principles calculations with excitonic effects}

\author{H. M. Lawler}
\affiliation{Dept.\ of Physics, Univ.\ of Washington Seattle, WA 98195}

\author{J. J. Rehr}
\affiliation{Dept.\ of Physics, Univ.\ of Washington Seattle, WA 98195}

\author{F. Vila}
\affiliation{Dept.\ of Physics, Univ.\ of Washington Seattle, WA 98195}

\author{S. D. Dalosto}
\affiliation{Dept.\ of Physics, Univ.\ of Washington Seattle, WA 98195}
\affiliation{National Institute of Standards and Technology, Gaithersburg,
MD 20899}

\author{E. L. Shirley }
\affiliation{National Institute of Standards and Technology, Gaithersburg,
MD 20899}

\author{Z. H. Levine}
\affiliation{National Institute of Standards and Technology, Gaithersburg,
MD 20899}


\date{\today}

\begin{abstract}
A first principles approach is presented for calculations of
optical -- ultraviolet (UV) spectra including excitonic effects.
The approach is based on Bethe-Salpeter equation calculations using
the \textsc{NBSE} code combined with ground-state density-functional theory
calculations from the electronic structure code \textsc{ABINIT}. 
Test calculations for bulk Si are presented, and the approach is
illustrated with calculations of the optical spectra and birefringence
of $\alpha$-phase SiO$_2$ and the rutile and anatase phases of TiO$_2$.
 An interpretation of the strong birefringence in TiO$_2$ is presented.
\end{abstract}
\pacs {PACS }

\maketitle

\section{Introduction}
        As in many subfields in condensed matter physics, the application 
of density-functional theory (DFT)\cite{2,2.1,2.2} has significantly influenced 
studies of material optical properties.  
However, treatments based 
on Kohn-Sham dipole transitions alone only provide a  beginning for
modeling optical spectra, since a ground-state one-electron approach
cannot account for a range of excited-state and optical effects.
Thus quantitative calculations of optical spectra require
the integration of DFT with a number of many-body techniques. 
Local field corrections from the random phase approximation (RPA)
alone are inadequate to reproduce near-gap optical spectra,
because they do not take into account self-energy and excitonic effects.
Consequently a quantitative theory and interpretation
must also include quasi-particle self-energy
effects based e.g., on the {\it GW}-method,\cite{3,3.1}
and particle-hole interactions based on the Bethe-Salpeter equation
(BSE). \cite{Hanke75} The success of the BSE method has effectively settled
the long-running discussion of the origin of many low energy spectral
features. \cite{1,tiago06}
For example, for the rutile phase of TiO$_2$, early studies could only
speculate about the relative importance of excitonic effects versus
band structure topology in the dominant threshold absorption,\cite{13}
whereas we show here that these features are dominated by  excitonic
effects.

While the BSE is often necessary for accurate calculations of optical
spectra, the approach has had limited applications, largely due to
a lack of efficient computational methods for general systems.
To address this need, we have developed an approach 
based on an interface, here dubbed \textsc{AI2NBSE},
between the BSE code \textsc{NBSE} developed at National Institute of Standards and Technology (NIST) \cite{nbsecode} and the general purpose DFT
electronic structure code \textsc{ABINIT}.\cite{4}
One of the objectives of this work 
is to improve the availability of BSE codes.
Our interface also provides a comparison to the BSE
codes \textsc{EXC}\cite{10.1}, and \textsc{EXC!ITING}, \cite{Exciting02}
and other recent codes, \cite{Schmidt03}
and also to approaches which model electron-hole interactions within
time-dependent density-functional theory (TDDFT).\cite{1.1}
The latter calculations are usually simpler than BSE ones,
but in practice are limited by an incomplete knowledge of the 
exchange-correlation functional and the neglect of damping effects. 
The overall strategy of our interface differs from
these approaches in several respects.
For example our \textsc{AI2NBSE}
interface achieves efficiency in the BSE calculations through
the use of the Hybertsen-Levine-Louie\cite{10,10.2}
dielectric screening, and can
also treat finite momentum transfer. Also, the interface requires 
only generic input, and thus can be adapted to other ground-state
and BSE codes.

The interface is tested on bulk Si, yielding results in good agreement
with other approaches. In particular we find that the calculations based
on \textsc{ABINIT} are in excellent agreement with those from the 
optimal basis function (\textsc{OBF}) code.\cite{nbsecode}
As initial applications, we report optical spectra and anisotropic
optical properties of the common rutile and anatase phases
of TiO$_2$, and for the $\alpha$-phase of SiO$_2$. These are
important materials for many applications.\cite{Mikami00,Shirley204}
However, we are not aware of earlier calculations for TiO$_2$ that include
electron-hole interactions, which are needed to reproduce their rich
dependence on polarization and phase.

In the remainder of this paper we briefly summarize the key formulas
describing excitonic effects
and other features of \textsc{NBSE}.  We then briefly describe our integration
of \textsc{ABINIT} and \textsc{NBSE}, using bulk Si as an illustrative test
case, with some further details in an Appendix. Finally, calculations
are presented for the birefringent spectra of the $\alpha$-phase SiO$_2$,
and similarly for the rutile and anatase phases of TiO$_2$, followed
by a summary and conclusions.


\section{BSE and Excitonic Effects}

\subsection{The Exciton Secular Equation}
In this paper we only briefly summarize the BSE formalism, following the
notation and theoretical developments of Shirley {\it et al}.\cite{shirley04}
Formally the BSE provides a complete theory for optical spectra
through a hierarchy of
equations derived from the
two-particle Green's function.\cite{6,6.1}
 However with certain approximations to the electron-hole interaction,
the BSE can be reduced to
an eigenvalue problem of an effective particle-hole Hamiltonian\cite{7}
\begin{equation}
 H |f\rangle = [H_{1e} + H_{2e}] |f\rangle = \Omega_f |f\rangle,
\end{equation}
where the eigenstates $|f\rangle$ are given by a superposition 
of particle-hole basis states\cite{8} $|nn'{\bf k(q)}\rangle$, {\it i.e.},
\begin{eqnarray}
&& |f\rangle = \sum_{nn'{\bf k}} \psi_f (nn'{\bf k(q)})\, |nn'{\bf k(q)}\rangle 
 \nonumber \\
&& |nn'{\bf k(q)}\rangle \equiv
a^{}_{n{\bf k}}a^{\dagger}_{n'{\bf k}+{q}}|0\rangle,
\end{eqnarray}
and throughout this paper we use atomic units ($e$=$\hbar$=$m$=1).
Here $\psi_f (nn'{\bf k(q)})$ is the amplitude of a given
particle-hole (or excitonic) state with Bloch crystal momentum index
$\bf k$ and
momentum transfer ${\bf q}$, the index $n$ runs over all occupied 
valence bands, the index $n'$ runs over unoccupied bands,
and $|0\rangle$ denotes the many-body ground-state with energy
$\Omega_0=0$.  For optical absorption ${\bf q}$ is
usually negligible. However for inelastic x-ray scattering and
other spectroscopies and for computational purposes, it is desirable to
retain the explicit momentum-transfer dependence.
The single-particle contribution to the Hamiltonian for a particle-hole
pair is diagonal in the quasi-particle/hole basis, so that one has
\begin{equation}
H_{1e} | nn'{\bf k(q)}\rangle  =
(E_{n'{\bf k}+{{\bf q}}}-E_{n{\bf k}})\, | nn'{\bf k(q)}\rangle , 
\end{equation}
where the quasi-particle energies $E_{n{\bf k}}$
are Kohn-Sham eigenvalues $\varepsilon_{n{\bf k}}$
plus quasiparticle self-energy corrections
\begin{equation}
E_{n{\bf k}} = \varepsilon_{n{\bf k}} + \Sigma_{n{\bf k}}.
\end{equation}
Here $\Sigma_{n{\bf k}}$ is the self-energy calculated in the {\it GW}
approximation for which efficient approximations have been
developed.\cite{shirley04,soininen03}

The electron-hole interaction contribution includes both ``direct" and 
``exchange" couplings $H_{2e}  = V_{D} + V_{X}$, 
\begin{eqnarray}
 && H_{2e} | nn'{\bf k(q)}\rangle = 
 \sum_{n'' n''' {\bf k}}
\left [V_D(nn'{\bf k},n'' n''' {\bf k'};{\bf q})\right. \nonumber \\
 &&\ \ + \left. V_X(nn'{\bf k},n'' n''' {\bf k'};{\bf q})\right] \,
| n''n'''{\bf k'(q)}\rangle,
\end{eqnarray}
with matrix elements defined explicitly below.
     Once the BSE secular equation is solved, the optical constants may be
obtained formally using a Fermi golden rule expression in terms of the
excitonic final states coupled to the current-like operator $J'_{\mu}$.
 In the \textsc{NBSE} code, however, these properties are calculated using
resolvent techniques.
In particular the imaginary part of the dielectric tensor
is given in terms of resolvents
\begin{eqnarray}
{\rm Im}\, \epsilon_{\mu\nu}(\omega) &=&
-\, 4 \pi\, {\rm Im}\, \left[ \langle 0| J'_{\mu}
  {[\omega-H+i\eta]}^{-1} J'_{\nu} |0\rangle \right. \nonumber \\
&& \ \ \ \  -\left. \langle 0| J'_{\nu} {[-\omega-H-i\eta]}^{-1} J'_{\mu}
|0\rangle\right].
\end{eqnarray}
In terms of the particle-hole states, the current-like operator coupling to
the ground-state is given approximately by
\begin{equation}
J'_{\mu} |0\rangle \approx \sum_{nn'{\bf k}} |nn'{\bf k(q)}\rangle
 \frac {  \langle \psi_{n'{\bf k}+{\bf q}} | J_{\mu} | \psi_{n{\bf k}}\rangle}
       {\varepsilon_{n'{\bf k}+{{\bf q}}}-\varepsilon_{n{\bf k}}},
\end{equation}
where $J_{\mu}$ is the $\mu$-th component of the current operator and
$\psi_{n{\bf k}}$ are approximated as Kohn-Sham one-particle states.
For small $q$, the matrix elements of $J_{\mu}$ are approximated by
\begin{eqnarray}
 &&\frac{1}{\omega} \langle \psi_{n'{\bf k}+{\bf q}} | J_{\mu}
| \psi_{n{\bf k}}\rangle
   \nonumber \\
&& \approx \left(
   \frac {\varepsilon_{n' {\bf k}+{\bf q} }-\varepsilon_{n{\bf k}} }
         { \omega q_{\mu}  } \right)
  \langle \psi_{n'{\bf k}+{\bf q}} | e^{i{\bf q}\cdot{\bf r}} |
           \psi_{n{\bf k}}\rangle \nonumber \\
&& \approx \frac{1}{q_{\mu}} \langle \psi_{n'{\bf k}+{\bf q} } |
                e^{ i {\bf q}\cdot{\bf r}} | \psi_{n{\bf k}}\rangle.
\end{eqnarray}
where ${\bf q}=q\hat\mu$. In the \textsc{NBSE} code these resolvents in the above
expressions for the dielectric tensor are calculated using an efficient
iterative Lanczos algorithm. \cite{haydock80}


\subsection{Electron-hole Interaction}

        The interaction kernel $H_{2e}$ of Eq.\ (5) accounts for two processes
which scatter an electron from band $n'$ to band $n'''$,
and a hole from band $n$ to band $n''$.  The first is the attractive
direct screened Coulomb interaction between the electron and the hole,
and the second is the repulsive unscreened exchange interaction.
Each of these contributions can be written in terms of two-particle
integrals between electron and hole Kohn-Sham orbitals
$\psi_{n{\bf k}}({\bf x})$ and $\psi_{n'{\bf k}+{\bf q}}({\bf x})$.
 Matrix elements of the direct term are given by
\begin{eqnarray}
&&V_{D} (nn'{\bf k},n'' n''' {\bf k'};{\bf q})
 \approx  -\int d^3x\, \psi^*_{n'''{\bf k'+q} } ({\bf x}) 
                         \psi^{}_{n'  {\bf k+q}  } ({\bf x})   \nonumber\\
  &&\times   \int d^3 x'\,  \psi^*_{n  {\bf k}    } ({\bf x'})
                            \psi^{}_{n'' {\bf k'}   } ({\bf x'})
                   W({\bf x,x'};\omega=0).
\end{eqnarray}
Here the electron-hole interaction $W({\bf x},{\bf x'};\omega)$
is the screened Coulomb attraction as mentioned above,
{\it i.e.}, $W=\epsilon^{-1}(\omega) V$,
which in an exact theory should include the frequency dependence of the
dielectric response. 
However the \textsc{NBSE} code
approximates the screening with the static, spatially dependent
Hybertsen-Levine-Louie dielectric function.\cite{10,10.2}  
This model maps the local density in real solids to the
density dependence in a homogeneous system.  
These calculations require the ground-state density, whose
Fourier coefficients are 
\begin{equation}
\rho({\bf G})=\frac {2}{N}
\sum_{n{\bf k},{\bf G}'} f_{n{\bf k}}\psi^{}_{n{\bf k}}({\bf G}+{\bf G}')
\psi^{*}_{n{\bf k}}({\bf G}'),
\end{equation}
where $f_{n{\bf k}} =\theta(\varepsilon_F- \varepsilon_{n{\bf k}})$
is the occupation of state ${n{\bf k}}$,
$N$ is the number of unit cells in the ABINIT calculation, which
is the same as the number of ${\bf k}$ points in the calculation,
and spin degeneracy is assumed.
Here $\psi_{n{\bf k}}({\bf G})$ are the Fourier coefficients of
the Bloch wave expansion\cite{10.2} in reciprocal lattice vectors
\begin{equation}
\psi_{n{\bf k}}({\bf x}) =
 e^{i{\bf k\cdot x}}\, \sum_{\bf G}\, \psi_{n{\bf k}}({\bf G})\,
 e^{i{\bf G}\cdot {\bf x}}.
\end{equation}
Because the dielectric function is modeled locally,
the exciton amplitudes
must be transformed from the Bloch basis
to a local basis with
coordinates ${\bf x}$ and ${\bf x'}$.
These local coordinates can 
be considered as transform analogs of the band and wave-vector indices
${n}$,  ${n'}$, and {\bf k}.\cite{5}

Similarly matrix elements of the 
exchange term are
\begin{eqnarray}
&&V_{X} (nn'{\bf k},n'' n''' {\bf k'};{\bf q})
 \approx +2 \int d^3x\, \psi^*_{n''' {\bf k'+q} } ({\bf x})
                        \psi^{}_{n'' {\bf k'}   } ({\bf x}) \nonumber \\
 &&\times \int d^3 x'\, \psi^*_{n   {\bf k}    } ({\bf x'})
                        \psi^{}_{n'  {\bf k +q} } ({\bf x'})
                       {1}{|{\bf x}-{\bf x'}|}.
\end{eqnarray}
As noted, for example by Hybertsen and Louie, the
exchange term in the BSE should be unscreened.

The matrix dimension of the
electron-hole interaction $H_{2e}$ is generally very large. For example,
for the calculations for SiO$_2$ presented below, 216 $\bf k$ points
are sampled, and there are 24 doubly degenerate valence (hole) states
and 26 conduction (electron) states. For this case, there are over
$10^5$ electron-hole pairs, each requiring a representation in both
periodic and local bases.


\subsection{ {AI2NBSE} Interface}

The above theory has been implemented in a modular code which
uses the output of ground-state density-functional theory calculations
from the electronic structure code \textsc{ABINIT} as input to
Bethe-Salpeter equation calculations using the \textsc{NBSE} code. The
interface serves as a driver for both \textsc{ABINIT} and \textsc{NBSE},
starting from a single input file, and also constructs the various physical
quantities and arrays needed in the calculations.  No explicit
changes in the structure or coding of either \textsc{ABINIT}
or \textsc{NBSE} are used. Both the interface and documentation are
available from the authors. \cite{sorini07}
Additional details are given in the Appendix.

\section{ Optical Spectra of Silicon}

 As a quantitative test, illustrative results from \textsc{AI2NBSE}
for bulk Si are presented in Fig.\ 1.
The two spectra compared are each calculations of the
imaginary part of the dielectric function for bulk Si using \textsc{NBSE}: 
in one case the ground-state quantities above are calculated with the
optimal basis function (\textsc{OBF}) 
pseudopotential, plane-wave code \cite{nbsecode} --
for which \textsc{NBSE} was originally designed -- and in the other case the
same quantities are calculated from \textsc{ABINIT}.  In both calculations
twenty valence and conduction bands were included with 
an eight Hartree cutoff criteria for the vectors {\bf G}.
The close quantitative agreement between \textsc{OBF} and \textsc{ABINIT}
thus serve as a quality check on the various theoretical and algorithmic
approximations used in our interface.

\def\Ref2{{Ref.~\onlinecite{Reining07}}}

\begin{figure}
\includegraphics[scale =.60, clip]{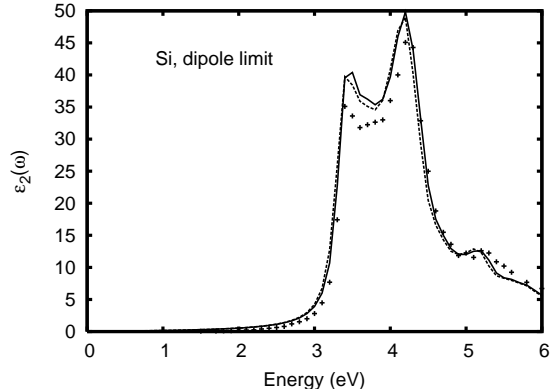}
\caption{Imaginary part of the dielectric function for Si using \textsc{NBSE}
and two different ground-state inputs; one set of inputs is from 
\textsc{OBF} (dashed line) and the other from \textsc{ABINIT} (solid line). 
For comparison the experimental spectrum is also plotted (crosses).}
\end{figure}

 Our \textsc{AI2NBSE} interface can also be applied to finite momentum transfer
calculations. For example,  Fig.\ 2 illustrates the spectrum for a
momentum transfer of magnitude $ q= 0.8$ $a_0^{-1}$, where $a_0$
is the Bohr radius.
 Results from a recent 
TDDFT calculation are also plotted.\cite{Reining07}
\begin{figure}
\includegraphics[scale =.60, clip]{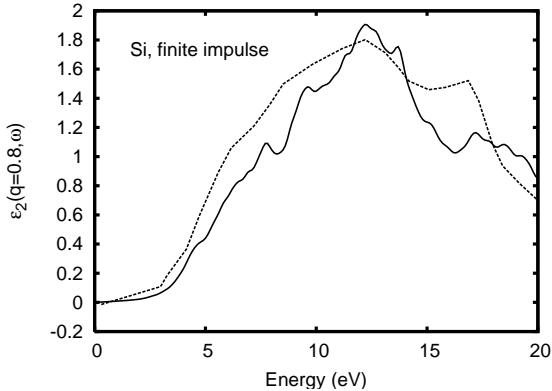}
\caption{Calculated $\epsilon_2(\bf{q},\omega)$ for Si with 
impulse magnitude of
$q = 0.8$ $a_0^{-1}$  
along the $[1,1,1]$ direction.
 \textsc{AI2NBSE}'s result (solid line), and for comparison the result
from a recent  TDDFT calculation (dashed line) are plotted.\cite{Reining07} }
\end{figure}

\def\Ref1{{Ref.~\onlinecite{8}}}

\begin{figure}
\includegraphics[scale=.6,clip]{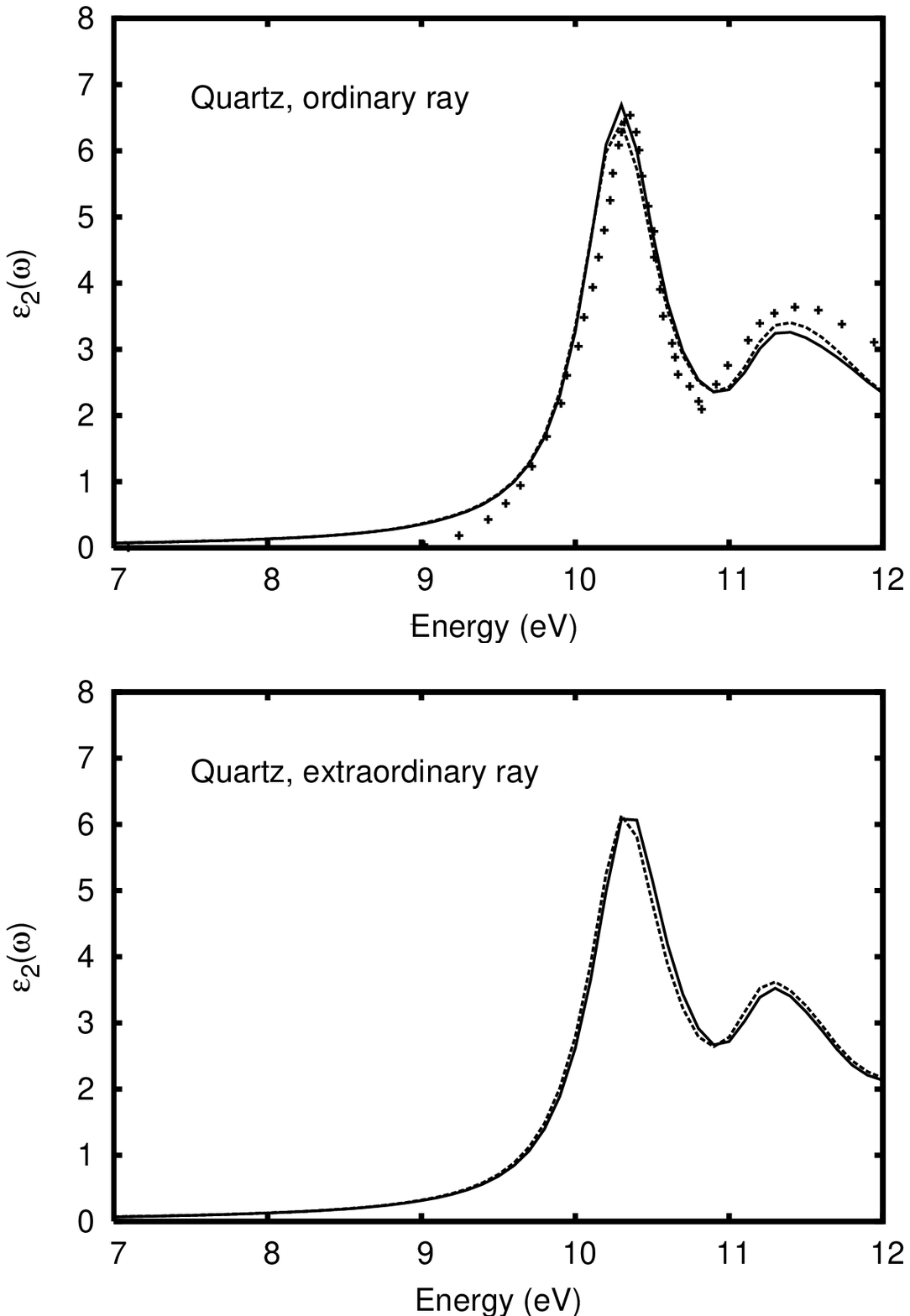}
\caption{Imaginary part of the dielectric function for SiO$_2$ with in-plane 
polarization (top), and out-of-plane polarization (bottom).  
Theoretical spectra from \textsc{ABINIT} and \textsc{OBF} (solid line and dashed line) spectra are plotted for
both polarizations.  The calculated spectra 
use a 28 Hartree plane-wave cutoff and included a 216-point Brillouin zone 
sampling and the 26 lowest-lying, degenerate bands above the Fermi level. 
The experimental spectrum, reproduced from 
\Ref1 is also shown (crosses) for the ordinary ray.}
\end{figure}

\section{Optical Spectra of $\alpha$-quartz}
    The excitonic character of the quartz optical spectra had been
recognized even before first-principles investigations 
became possible.\cite{Laugh,Sok}
The large band gap and small static dielectric constant of $\alpha$-quartz--and 
the prominent low-energy features in the spectrum--suggested 
a strong excitonic role in the optical properties, which was subsequently 
confirmed by first-principles calculations. \cite {8} 
Our calculations, like those of Ref.\ \onlinecite {8}, reproduce
the measured spectra in detail, with excellent agreement in oscillator strength 
and calculated feature positions corresponding with those measured to within
tenths of an eV. The results can be seen in Fig.\ 3. 

The calculations of Ref. \onlinecite {8} were performed for photon
polarization
within the hexagonal plane ({\it i.e.}, the ``ordinary ray"),
noting that this is the most commonly measured spectrum.
This material is however uniaxial, possessing an independent optic
axis normal to the hexagonal plane.
	We have also carried out calculations for photon 
polarization perpendicular to the hexagonal plane ({\it i.e.}, the 
``extraordinary ray").  As illustrated in Fig.\ 3, quartz exhibits
some anisotropy, but it is less pronounced than for rutile.
This result is consistent with the known 
role of the SiO$_4$ tetrahedron versus the anisotropic crystal structure, 
and the structural phase insensitivity of
the optical functions in quartz. \cite{Laugh}

While birefringence in quartz is a well-studied effect, it is 
small relative to TiO$_2$ crystals discussed in the the next section.
Below the interband transition energies, 
but above the lattice response, we 
find static indices of refraction
of 1.52 for the ordinary polarization, and 1.53 for the
extraordinary at 0.7 eV, in precise agreement with the measured
values.\cite{Sosman27}

\section{Optical Spectra of Rutile and Anatase}

      As one of the simplest transition-metal oxides, TiO$_2$
exhibits a variety of natural crystal structures and presents a 
fundamentally interesting system for first-principles electronic 
structure methods.  This material is also an important component in
various ultraviolet (UV) applications.  It has been demonstrated 
that {\it ab initio} methods can describe various physical 
properties of TiO$_2$. \cite{Mikami00,Shirley204}
There have been a few {\it ab initio} studies of  
UV dielectric spectra for rutile and anatase phases.
Glassford and Chelikowsky\cite{11}
reported calculations for the rutile phase using
a plane-wave pseudopotential approach; Mo and Ching\cite{Mo} used a
linear-combination-of-atomic orbital method for both 
phases (and brookite); and Asahi {\it et al.}\cite{12}
studied the anatase phase with 
a linearized augmented plane-wave method. However, none of these studies
included excitonic effects.  
A theoretical treatment that includes excitonic states 
may contribute to understanding its optical properties.
Toward this end, we present calculations of the spectra 
of TiO$_2$, with polarization dependence, 
for both the rutile and anatase phases using our \textsc{AI2NBSE}
interface.

Each of the two TiO$_2$ phases is tetragonal, and Mo and Ching\cite{Mo}
and Fahmi {\it et al.} \cite{Fahmi93} have reviewed their structural properties including
crystal structures, space groups, and differences in bond lengths and
angles.
These two crystal 
structures can be considered as arrangements of slightly distorted 
oxygen octahedral elements with a titanium atom at the center of each, 
so that each titanium has an oxygen coordination of six, and each 
oxygen has a titanium coordination of three.
The relationship between the 
two structures has been described in terms of varied orientation among the 
octahedral chains.
The two polymorphs studied here can be generated with 
six-atom unit cells corresponding to two TiO$_2$ units. 
Each unit possesses two inequivalent bonds of ``apical'' and
``equatorial'' character, such that 
each titanium sees two apical and four equatorial bonds, while
and each oxygen sees one apical and two equatorial bonds.

     Structural similarities in rutile and anatase lead to 
similarities in their electronic structure.  
In Fig.\ 4 we plot the \textsc{ABINIT}-calculated 
densities of states for 
for the two materials.
The local-density approximation (LDA) gives  
the uppermost valence-band width of about 5 eV for anatase and 6 eV for
rutile, and this band is regarded to be dominated by O 2$p$ character.
The anatase LDA band gap is greater than that of rutile by 
about 0.2 eV, corresponding with the measured difference.\cite{Reyes-Coronado}
Because  the oxygen bonding environment 
is planar in both phases, there is a  decomposition of the O 
2$p$ into $p_{\pi}$ 
and $p_{\sigma}$ states.\cite{12,Sorantin}

The first conduction band is dominated by Ti 3$d$ character,
with the lower half regarded as $t_{2g}$-like, and the upper half 
regarded as $e_g$-like.\cite{12,Sorantin}  
As can be seen in Fig.\ 4 and
from the band structures,\cite{Mikami00}
these sub-bands are reasonably resolved energetically.
Our calculations were performed with semicore states treated as core using
Troullier-Martins-type pseudopotentials,\cite{TroulMart,11} and as valence
using Teter-type pseudopotentials.\cite{Teter,Mikami00}
 While the semicore states have spatial localizations comparable to 
oxygen 2$p$ states, their energies are many eV below the Fermi level.
Our calculations suggest that the optical and UV spectra
are not highly sensitive to the treatment of Ti semicore states,
and that the primary source of the
discrepancy between theoretical and experimental spectra
is the neglect of the excitonic effects in the low-energy region.



\begin{figure}
\includegraphics[scale=.6,clip]{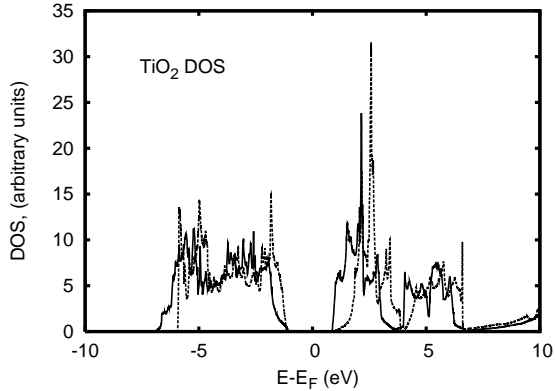}
\caption{Density of states $\rho(E)$ with respect to the Fermi energy
$E_F$ for the uppermost valence and lowest conduction 
bands of TiO$_2$ in the rutile (solid line) and anatase (dashed line) phases.
No band gap corrections
are included.  The LDA gap is calculated to be 2.0 eV
for rutile and 2.2 eV for anatase.}
\end{figure}
 

 \def\Ref3{{Ref.~\onlinecite{13}}}
\begin{figure}
\includegraphics[scale=.6,clip]{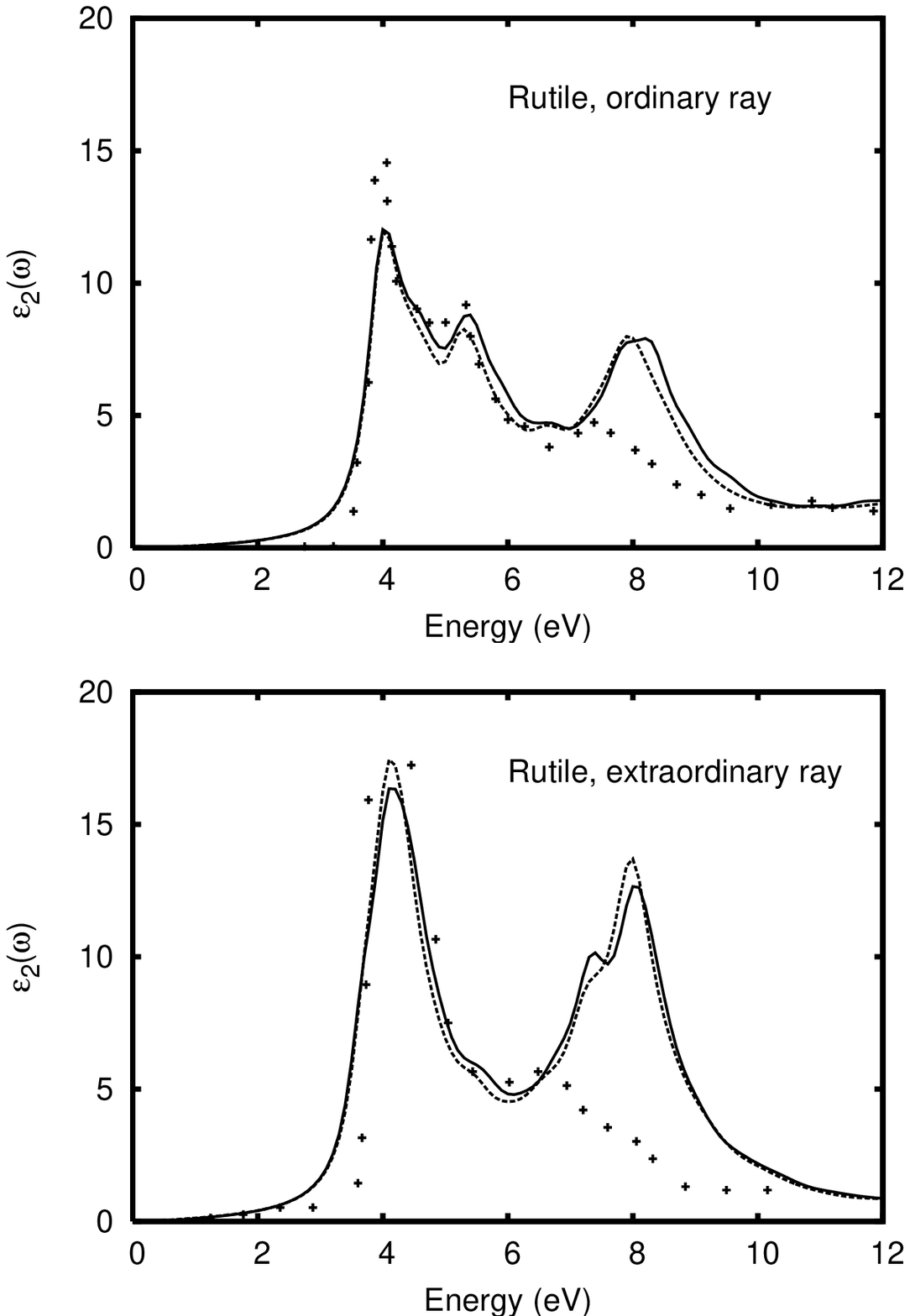}
\caption{Imaginary part of the dielectric function for rutile with in-plane 
polarization (top) and out-of-plane polarization (bottom).  
Theoretical from \textsc{ABINIT} and \textsc{OBF} 
(solid line and dashed line) spectra are plotted for
both polarizations.  The calculated spectra 
use a 32 Hartree plane-wave cutoff and included a 216-point Brillouin zone 
sampling and the 26 lowest-lying, degenerate bands above the Fermi level.  
For comparison the experimental spectrum from \Ref3 (crosses) is
also plotted.
}
\end{figure}


\subsection{Rutile Phase}        

The spectra calculated for rutile with the \textsc{AI2NBSE} interface
(Fig.\ 5) demonstrate significant excitonic effects, as can
be seen by comparison with calculations (Fig.\ 6) which
neglect electron-hole interactions and are calculated within the RPA.
One signature of the excitonic effects
is the sharp onset in our BSE calculations, 
which better reproduces experimental spectra.
\begin{figure}
\includegraphics[scale=.6,clip]{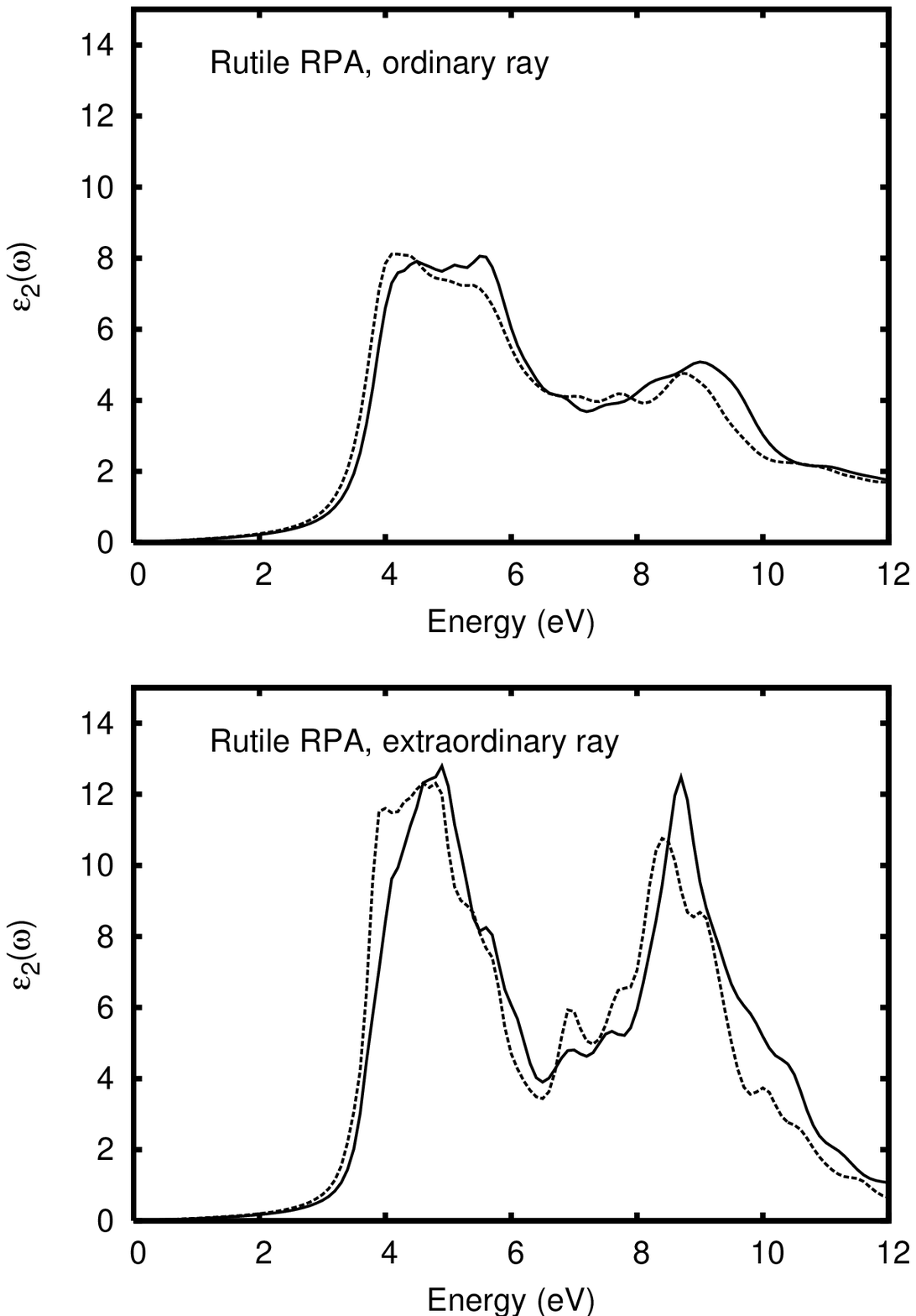}
\caption{Imaginary part of the dielectric function for rutile with in-plane
polarization (top) and out-of-plane polarization (bottom) calculated
without core-hole interactions ($V_D=0$).
 Two spectra are plotted corresponding to a treatment of semicore states
with Teter-type Ti pseudopotentials (solid line),
and with Troullier-Martins pseudopotentials where the semicore states
are pseudized (dashed line).}
\end{figure}
\begin{figure}[t]
\includegraphics[scale=.6,clip]{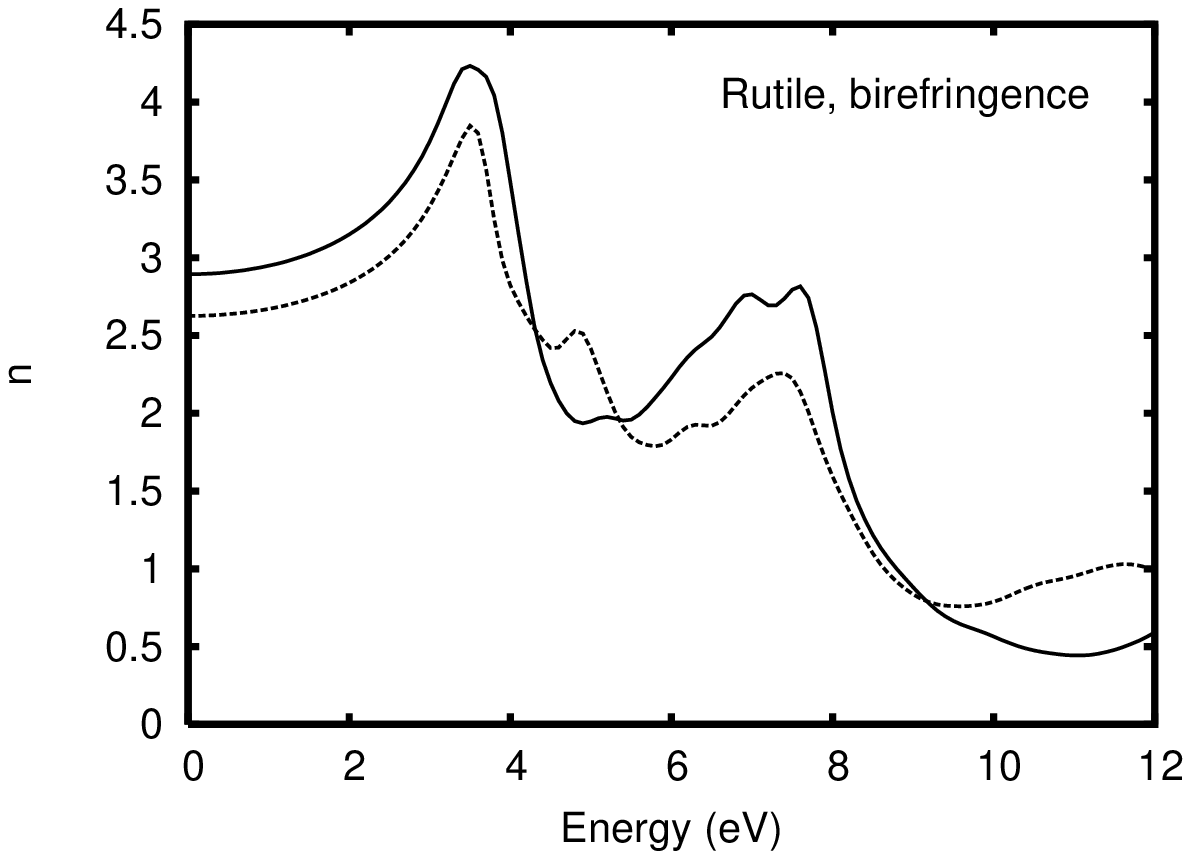}
\caption{Ordinary (dashed line) and extraordinary (solid line) 
indices of refraction $n$ of rutile-TiO$_2$ as a function of photon energy.  
At 589 nm, which is below the band gap but well above the lattice resonances, 
the calculated difference between the two indices is 0.30
while the measured value is 0.29.\cite{13}}
\end{figure}
        Our results for the polarization dependence in TiO$_2$
show a much more prominent anisotropic optical response than in SiO$_2$.
 This results in strong birefringence, as seen in Fig.\ 7.
Our calculated static indices of refraction differ by 0.30,
in agreement with the observed birefringence.\cite{13}

Two pronounced anisotropic features in
our calculations are consistent with experimental spectra.\cite{13,Jell}
 First, the primary onset absorption feature assumes 
a doublet structure for the ordinary ray which is not evident for the 
extraordinary ray.  These low-energy excitonic features
at 4.0 and 5.3 eV are clearly evident in our calculation 
and are expected to involve $t_{2g}$-like final states.\cite{Mo}
Second, the onset of absorption for the extraordinary ray is significantly 
stronger than for the ordinary.     For the extraordinary ray, a
singe feature is measured at 4.1 eV, which is stronger, broader,
and more symmetric than for the ordinary ray. 
However, a third anisotropic feature of our calculations, namely the much stronger 
absorption peak for the extraordinary ray above 6 eV, is not clearly
seen in the experimental reflectivity spectra.  It should be noted that the 
peak at 7.4 eV
is more prominent in the x-ray absorption spectra than in the reflectivity data, \cite{11,Fischer72}
and is expected to involve 
$e_g$-like conduction states.\cite{Mo}
To summarize,
our BSE calculation gives a stronger, less structured $t_{2g}$-like 
absorption band for the extraordinary ray than for the ordinary, in accord
with experiment. 

	
\subsection{Anatase Phase}

     Theoretically, both the valence-band and conduction-band densities
of states are similar to that of rutile, as shown in Fig.\ 4.   
However,
their experimental absorption spectra show more variation (Fig.\ 8.).
\begin{figure}
\includegraphics[scale=.6,clip]{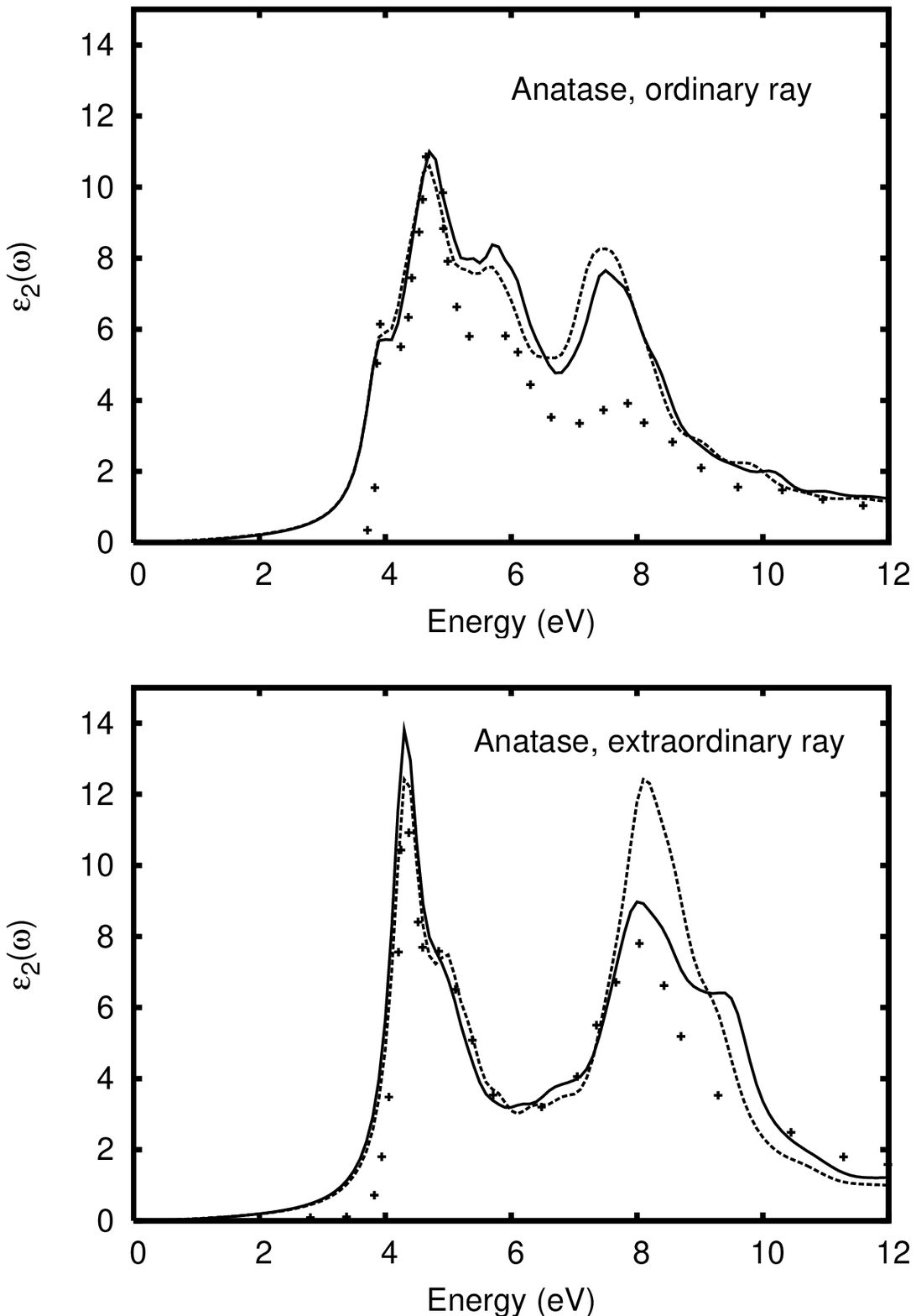}
\caption{Imaginary part of the dielectric function for anatase with in-plane 
polarization (top) and out-of-plane polarization (bottom).  
Theoretical spectra from \textsc{ABINIT} and 
\textsc{OBF} (solid and dashed lines) are plotted for
both polarizations.  The calculated spectra use a 32 Hartree plane-wave
cutoff and include a 216-point Brillouin zone 
sampling and the 26 
lowest-lying, degenerate bands above the Fermi level.  
For comparison the experimental spectrum reproduced from \Ref3 (crosses)
also plotted.}
\end{figure} 
      Strikingly, the measured spectrum of anatase 
for the ordinary ray (in-plane polarized)
shows a clear low-energy 
shoulder below the primary absorption feature.\cite{Hosaka,Jell} 
This feature is 
not seen in rutile, and calculations neglecting the electron-hole interactions 
in anatase do not clearly resolve it.\cite{12,Mo}  Aside from this low-energy 
feature, the spectral structure and anisotropy of anatase, 
both calculated and measured, have some similarities to that of rutile.
Thus features in the spectra have been 
analyzed,\cite{12,Hosaka} and conclusions regarding the relevant bands are 
reminiscent of results for rutile.\cite{11}  
     For the $e_g$-like absorption band, the experimental spectrum for
anatase does show a stronger absorption for the extraordinary 
ray,\cite{Hosaka} similar to 
what is calculated in both the rutile and anatase phases, but not seen in 
experimental rutile spectrum.
 

\subsection{Interpretation of Polarization Dependence }

     Molecular orbital interpretations of the TiO$_2$ spectra
have been reported widely \cite {Fischer72,Modrow03} and
provide a qualitative picture of the electronic structure in this system.  
Such analyses illustrate the hybridization between the Ti 3$d$ and 
O 2$p$ states.  However, immediately above 
and below the Fermi level, the hybridization has been 
shown to be weak.\cite{Sorantin,12} 
The calculations of Sorantin and Schwartz \cite{12}
demonstrate that the valence band for rutile just
below the Fermi level is primarily $\pi$-bonding  O 2$p$
character, while lower in energy is primarily $\sigma$-bonding O 2$p$
character.  Just above the Fermi level, the states are, in energetic order, primarily 
Ti $t_{2g}$ and $e_g$.

A molecular-orbital analysis, considered in conjunction with measured x-ray
spectra, allowed an empirical identification of features in the optical
spectra of Harbeke and Cardona\cite{13} with specific electronic
transitions.\cite{Fischer72}  Although selection rules, 
inferred from the approximate atomic states, were
invoked in that work, no effort was made to analyze the polarization
dependence of the selection rules. 
Some of the most prominent features could not be identified at all, and 
were 
attributed to the unaccounted excitonic effects.  Assignments of
the same experimental features 
were made, for ordinary and extraordinary rays independently,
from a band-structure perspective by Glassford and Chelikowsky.\cite{11}   But the 
calculated spectra again did not treat the excitonic interactions, and hence
did not reproduce the experimental features to the level of agreement
reported here.  Also, local electronic structure was not emphasized
in the peak assignment analysis.  

Perhaps the most conspicuous anisotropic characteristic in the rutile spectra 
is the stronger onset absorption for the extraordinary ray.  We observe that the
first eV below the Fermi level is dominated by the O 2$p_{\pi}$ states, and further, that the 
oxygen bonding planes are all defined by in-hexagonal-plane normals.  This implies that, 
for some in-plane polarization, 
there is a single dipole allowed channel from one of the two 
$\pi$-state orientations (corresponding to the two oxygen-bonding planes), 
and the three $t_{2g}$ states, while for the extraordinary polarization, there are 
two dipole-allowed channels.
This is consistent with the calculated and measured stronger
threshold resonance for the extraordinary ray.
In connection with this point, we emphasize that while the $t_{2g}$ and
$e_g$ symmetry labels are approximate because of angular and bond-length 
distortions within the octahedra, the selection rules above are unmodified 
from the above even when the local, $D_{2h}$ symmetry group is considered, 
while the actual crystallographic group in rutile is of even higher 
symmetry.

\section {Summary and Conclusions}

In summary, 
we have developed a first principles approach for calculations
of various optical spectra, including finite momentum transfer in crystals.
The method combines ground state electronic
structure calculations from \textsc{ABINIT} with BSE calculations from
the \textsc{NBSE} code. The method is tested on  bulk Si, yielding 
results in good agreement with other methods.
Calculations are presented for the macroscopic dielectric spectra and
its orientation dependence in $\alpha$-quartz, rutile TiO$_2$, and anatase
TiO$_2$.
Our quartz spectra for the ordinary ray reproduce the strong excitonic 
character and are in good agreement with experiment and earlier theoretical
work. 
The anisotropy of the rutile and anatase phases
of TiO$_2$ are more significant than for quartz.  
The static indices of refraction for the two polarizations of rutile 
differ by more than 10\%, in agreement with experiment.  Also our calculated
absorption at low energies reproduces experiment better than previous
theoretical results that neglect excitonic effects. In particular our calculations
reveal an additional low-energy feature in anatase also found in experiment.
Also we are able to interpret the anisotropy in the threshold behavior 
for rutile in terms of $\pi$ to $t_{2g}$ selection rules.

\acknowledgments

This work was supported by DOE Grant DE-FG03-97ER45623 (JJR,HML,SDD),
by NIST Grant 70 NANB7H6141 (SDD), and was
facilitated by the DOE Computational Materials Science Network.
M.\ Prange, J. Kas and A.\ Sorini assisted with testing the interface.
We also
thank L.\ Reining, V.\ Olevano, S.\ Ismail-Beigi, and especially X.\ Gonze
and the \textsc{ABINIT} development group for helpful discussions.


\appendix*

\section{\textsc{AI2NBSE} Interface}

 In order to calculate the optical spectra, the \textsc{AI2NBSE} 
interface first obtains Kohn-Sham energies and wave functions from 
the self-consistent ground-state electronic structure code \textsc{ABINIT}.
The single-particle eigenenergies are modified to include self-energy
corrections according to the {\it GW} approximation.  Subsequently the interface
constructs several quantities needed for the \textsc{NBSE} calculations.
These include current--operator matrix elements between Kohn-Sham states and
the ground-state charge density for calculating Hybertsen-Levine-Louie
screening.  Thus our interface does not take advantage of \textsc{ABINIT}'s
dielectric function capability.

Typical \textsc{AI2NBSE} calculations are divided into four modular stages and
require a single input file which contains all parameters needed to
define both the system and its ground-state and excited state one-electron
properties.
Briefly the modular operations are as follows after the input file
is read and stored:

\noindent 1) \textsc{ABINIT} {\it calculation}:
\textsc{ABINIT} inputs are generated and \textsc{ABINIT} is run.
 These calculations supply the Kohn-Sham eigenvalues
$\varepsilon_{n{\bf k}}$
and eigenfunctions $\psi_{n{\bf k}}$ both
for the occupied and unoccupied states.
 Currently the interface uses a standard serial distribution
of \textsc{ABINIT}. However, calculations for large systems can still be performed
by means of consecutive runs which are automatically set up by the interface,
with minimal input or intervention from a user.

\noindent 2) {\it Density components}: Fourier components of the
ground-state density $\rho({\bf G})$ are generated
using the eigenfunctions from stage 1.


\noindent 3) {\it Dipole matrix elements}: Dipole matrix elements
in Eq.\ (8) are calculated, and the eigenfunctions are converted to
the format used by \textsc{NBSE}.

\noindent 4) \textsc{NBSE} {\it calculation}:
All quantities required for the Bethe-Salpeter
calculation are collected and \textsc{NBSE} is run.
The output includes various optical spectra and optical constants.

\end{document}